\documentclass[12pt]{article}
\usepackage{amsmath,amssymb,graphicx,hyperref,natbib}
\usepackage[margin=1in]{geometry}
\usepackage{algorithm}
\usepackage{algpseudocode}
\usepackage{tikz}
\usepackage{float}
\usepackage{booktabs}
\usepackage{longtable}
\usepackage{mdframed}
\usepackage{booktabs}
\usepackage{enumitem}
\usepackage{hyperref}

\hypersetup{
  pdfauthor = {Michael S. Harré},
  pdftitle  = {From Firms to Computation: AI Governance and the Evolution of Institutions}
}

\title{From Firms to Computation: \\
AI Governance and the Evolution of Institutions}
\author{Michael S. Harr\'e \\
\small School of Computer Science \\
\small Centre for AI Trust and Governance \\
\small University of Sydney\\
\small \texttt{michael.harre@sydney.edu.au}}
\date{\today}

\begin{document}
\maketitle

\begin{abstract}
The integration of agential artificial intelligence into socioeconomic systems requires us to reexamine the evolutionary processes that describe changes in our economic institutions. This article synthesizes three frameworks: multi-level selection theory, Aoki's view of firms as computational processes, and Ostrom's design principles for robust institutions. We develop a framework where selection operates concurrently across organizational levels, firms implement distributed inference via game-theoretic architectures, and Ostrom-style rules evolve as alignment mechanisms that address AI-related risks. This synthesis yields a multi-level Price equation expressed over nested games, providing quantitative metrics for how selection and governance co-determine economic outcomes. We examine connections to Acemoglu's work on inclusive institutions, analyze how institutional structures shape AI deployment, and demonstrate the framework's explanatory power via case studies. We conclude by proposing a set of design principles that operationalize alignment between humans and AI across institutional layers, enabling scalable, adaptive, and inclusive governance of agential AI systems. We conclude with practical policy recommendations and further research to extend these principles into real-world implementation.
\end{abstract}

\medskip
\noindent\textbf{Keywords:} AI governance; Institutional economics; Multi-level selection; Polycentric policy; Agential AI;\newline 
{\bf AI-assistance declaration:} Drafting benefited from OpenAI ChatGPT-4o, o3 and Anthropic Claude-3. The author verified all content.

\hypersetup{linktoc=all}
\tableofcontents
\clearpage  

\section{Introduction}

The integration of artificial intelligence (AI) into economic organizations represents a profound transformation, raising fundamental questions about institutional architecture, governance, and evolutionary dynamics. We focus specifically on {\it agential} artificial intelligence: AI that acts as an autonomous decision-making entity, capable of pursuing goals, contextually adapting its behavior, and influencing outcomes within institutional environments. As firms deploy these AIs our traditional frameworks for understanding organizational behavior and its evolution over time and through an evolving strategic state space requires significant extension. In many ways they act as computational economic agents that for modelling purposes can be thought of as having some degree of {\it autonomy} or {\it artificial agency}~\cite{rahwan2019machine,crandall2018cooperating}. See for example Wood~\cite{wood2021algorithmic} on the movement towards {\it algorithmic management} in which AIs ``... are deployed to automate workforce direction, evaluation and discipline.'' This article develops a framework that bridges evolutionary theory, institutional economics, and AI governance to address these new challenges.

Our approach integrates three foundational theoretical strands. First, multi-level selection (MLS) theory provides an evolutionary framework for understanding how selection operates simultaneously across multiple organizational levels—from individual agents to teams, divisions, and entire corporations \cite{Traulsen2006, Wilson2007}. Second, Masahiko Aoki's conceptualization of firms as endogenously evolving computational devices offers insights into how information processing occurs through game-theoretic interactions among organizational components \cite{Aoki2010}. Third, Elinor Ostrom's design principles for common-pool resource institutions provide empirically-grounded guidance for creating robust governance systems that balance autonomy with accountability \cite{Ostrom1990, Ostrom2005}.

\begin{mdframed}[linewidth=0.5pt, backgroundcolor=gray!10]
\textbf{Alignment Terminology}

Throughout this paper, we distinguish between two related concepts:

\textbf{Alignment operators} refer to high-level institutional mechanisms (such as Ostrom's design principles) that transform interaction structures to promote cooperative outcomes.

\textbf{Alignment parameters} are specific quantifiable elements ($\lambda$, $k_i$, etc.) that implement these operators within formal models, modifying payoff structures and incentives.
\end{mdframed}

Corporations face what we might call the agential-alignment problem (AAP)—ensuring that the actions of semi-autonomous agents (both human and algorithmic) serve organizational objectives rather than narrowly defined self-interest. Traditional single-agent, selfish evolutionary processes do not allow for evolution to act specifically on interactions between agents. In contrast, MLS provides a mechanism through which interaction patterns themselves become subject to evolutionary processes, helping explain how institutional rules that promote collective fitness emerge.

The organizational routines that encode operational knowledge within firms are analogous to genes in biological evolution \cite{Nelson1982, Becker2004}. These routines store institutional knowledge and evolve in response to selection pressures. However, unlike biological genes, routines operate within explicitly designed governance structures. The co-evolution of routines and formal institutional rules creates a complex dynamic wherein organizational learning, memory, and governance structures interact—a dynamic that becomes even more complex when AI systems enter the picture.

\subsection{From Biological to Institutional MLS}
Multi-level selection (MLS) theory extends Darwinian evolution to contexts where fitness is simultaneously determined at multiple organisational layers—genes, individuals, groups, and metapopulations~\cite{Okasha2006,eldakar2011eight}. Economic systems exhibit analogous hierarchies: employees nest within teams; teams within divisions; divisions within corporations; and firms within broader industrial and regulatory ecologies. Recent work applies MLS to explain the emergence of cooperative norms and institutional diversity in economic development~\cite{Wilson2007,BowlesGintis2011}.

While classical MLS theory has primarily focused on biological selection processes, its application to cultural and institutional evolution requires careful extension \cite{Henrich2004, Richerson2005}. Institutions represent a unique level of organization in human societies—one that explicitly encodes rules, norms, and governance structures that shape individual and group behaviors. Unlike genetic inheritance, institutional knowledge is transmitted both vertically (across generations) and horizontally (across organizations), creating complex evolutionary dynamics \cite{Boyd1985}. In Table~\ref{tab:Bio-Eco} we provide a mapping from the categorical elements of biological systems to that of institutional economics. The reference to ``bottlenecks'' and ``recursive solutions'' is a reference to the work in~\cite{muro2025emergence}, also see Section~\ref{sec:Inst-Genes-MLS} 

\begin{table}[h]
\centering
\label{tab:inheritance-mapping}
\begin{tabular}{@{}p{4cm}p{5cm}p{5cm}@{}}
\toprule
\textbf{Level, Role} & \textbf{Biological Systems} & \textbf{Institutional Economics} \\
\midrule
Individual unit & Gene or single cell & Person or firm \\
Expressed trait & Protein-coding expression & Strategic behavior or decision \\
Interaction structure & Regulatory gene networks & Institutions: rules, norms, legal frameworks \\
Inheritance mechanism & Genetic replication & Dual inheritance (agent learning \& social transmission) \\
Selection context & Organism-level fitness & Firm's success or governance performance \\
Evolutionary bottleneck & Combinatorial overload in gene regulation & Coordination failure, transaction cost explosion \\
Recursive solution & Non-coding regulatory DNA & Emergent governance: constitutional rules, meta-institutions \\
\bottomrule
\end{tabular}
\caption{Mapping MLS from Biology to Economics} \label{tab:Bio-Eco}
\end{table}

\subsection{The Computational Turn in Economic Institutions}
Traditional economic theory has conceived of institutions primarily as rule systems or incentive structures. However, the computational perspective, pioneered by \cite{Aoki2010}, reframes institutions as distributed information-processing architectures that coordinate decentralized decisions. This computational turn has gained new significance with the integration of AI systems as autonomous or semi-autonomous agents within economic organizations. 

Aoki's framework conceptualizes institutions as equilibrium states in societal games -- configurations that coordinate beliefs and behaviors through shared cognitive representations \cite{Aoki2001, Aoki2010}. This perspective aligns with recent advances in distributed computing and multi-agent systems, where coordination emerges from local interactions governed by shared protocols \cite{Jennings1996, Wooldridge2009}.

As firms increasingly deploy machine learning algorithms, the computational capacity of the organization itself undergoes qualitative transformation, raising fundamental questions about how selection pressures operate on hybrid human-AI organizations and how institutional rules might evolve to ensure alignment between algorithmic optimization and broader organizational objectives.

\subsection{Acemoglu's Institutional Economics and AI}
\cite{Acemoglu2012} has advanced our understanding of how institutional structures shape economic development through their effects on innovation, resource allocation, and power distribution. A key element of this work is the distinction between inclusive institutions that distribute economic opportunity broadly and extractive institutions that concentrate benefits among elites. This framework offers an important insight into the evolution and historical contingencies of successful nation states and also provides a perspective from which to consider how AI technologies might be deployed in ways that either reinforce existing inequalities or promote broader participation.

In recent work on AI, Acemoglu has focused primarily on AI as a factor of economic production rather than viewing it as an {\it agential} technology. This approach separates AI's economic impact into different channels: automation (AI replacing humans in tasks), productivity enhancement (AI making humans more productive at existing tasks), and the creation of new tasks \cite{Acemoglu2021harms}. Notably, this economic perspective differs from approaches that view AI systems as behaving as agents in their own right, as proposed by Rahwan et al~\cite{rahwan2019machine} in their ``machine behavior'' framework.

Acemoglu's perspective particularly emphasizes the economic niches that AI might occupy—replacing workers through ``so-so automation''\footnote{Technologies that {\it ... are good enough to be adopted, but not so good as to have a meaningful impact on productivity.}~\cite{Acemoglu2021harms}} with limited productivity gains, or alternatively complementing human labor in ways that augment capabilities and create new economic roles \cite{Acemoglu2021harms}. He argues that current AI development has overemphasized direct labor replacement rather than creating complementarities where AI supports human work. This relates to our multi-level selection framework, as different institutional arrangements may favor different patterns of AI development and critically its deployment into the economic workspace, leading to significant variations in economic and social outcomes.

The work of Acemoglu and Robinson emphasizes the path-dependent nature of institutional development and the critical role of power dynamics in determining which institutional arrangements prevail. Applied to AI governance, this suggests that early decisions about regulatory frameworks, corporate deployment patterns, and algorithmic design principles may have long-lasting consequences for the distribution of AI's economic benefits and risks \cite{Acemoglu2021harms}. Unlike Rahwan et al's approach~\cite{rahwan2019machine}, which studies the behavior of AI systems as entities that require scientific investigation in their own right, Acemoglu focuses on how institutional structures determine the development trajectory of AI applications and their subsequent economic impacts.

\subsection{Institutional Inheritance, Genetic Regulation, and Multi-Level Selection \label{sec:Inst-Genes-MLS}}

The institutional economics view highlights the importance of interaction-level structures -- rules, norms, and governance mechanisms -- that shape collective outcomes beyond individual incentives. This line of reasoning has a parallel in evolutionary theory, particularly in the frameworks of multi-level selection (MLS) and dual-inheritance theory (DIT). In this article we will propose a general analogy: just as complex regulatory architectures in biology evolve to coordinate interactions among genes and cells, institutions emerged to regulate agent interactions in economic systems. This is particularly notable as the transition from single-celled to multi-cellular organisms has been shown to be an evolutionary {\it phase transition} that resolved a computational optimisation problem~\cite{muro2025emergence}. Here we interpret the emergence of the {\it corporation}~\cite{Aoki2001,Aoki2010} or firm as a solution to an economic coordination problem that single economic agents needs to resolve when cooperation and coordination is needed to carry out complex economic processes.

\subsection{Purpose and Scope}
Our aim is threefold. First, we establish a generalized multi-level selection framework for evolutionary game dynamics in economic hierarchies, explicitly accounting for AI agents as distinct evolutionary units. Second, we show how Aoki's corporations-as-computers paradigm~\cite{Aoki2010} and Ostrom's {\it common-pool resource} (CPR) design principles~\cite{Ostrom1990} jointly function as endogenous alignment operators within this framework. Third, we integrate Acemoglu's insights on institutional development to analyze how power asymmetries and historical contingencies shape the evolution of AI governance regimes.

We pay special attention to the novel agential risks introduced when AI agents become strategic actors within corporate decision-making. By treating AI agents as replicating inferential agents subject to selection for predictive accuracy and profit, we locate AI safety within the broader MLS–institutional framework. Throughout, we ground this theoretical framework in empirical evidence from organisational studies, evolutionary economics, and emerging patterns in AI deployment.

\section{Theoretical Foundations \label{sec:Theo-Found}}
\subsection{Multi-Level Selection and Economic Organizations}

\subsubsection{Multi-Level Selection and the Price Equation}
The core intuition of multi-level selection is straightforward: selection can operate simultaneously at multiple organizational levels. At each tier of organization, variation in reproductive success that correlates with performance transmits selection effects upward through nested organizational levels. 

When selection operates across multiple levels, the direction and strength of evolution depends on the balance between within-group competition (which may favor selfish strategies) and between-group competition (which often favors cooperative strategies). This balance is shaped by institutional structures that modify incentives, information flows, and interaction patterns.

The Price equation \cite{Price1972,lehtonen2020price} provides a formal framework for understanding how selection operates across multiple levels. In its simplest form for economic organizations, the change in average organizational performance ($\Delta\overline{\pi}$) can be decomposed into~\cite{okasha2020price}:

\begin{equation}
\overline{w} \Delta\overline{\pi} = \underbrace{\text{Cov}_g(w^g,\pi^g)}_{\text{between-group}} \,\, + \,\, \underbrace{\mathbb{E}_g\left[\text{Cov}_i(w^g_i,\pi^g_i)\right]}_{\text{within-group}}
\label{eq:MLPrice}
\end{equation}

Here, $w_i^g$ and $\pi_i^g$ represent the fitness and genetic value, respectively, of individual $i$ in group $g$, $w^g$ and $\pi^g$ are the corresponding values for group $g$, while $\overline{w}$ and $\overline{\pi}$ are the mean fitness and genetic values across all groups. The first term captures selection between groups, where groups with higher performance tend to have higher fitness and thus contribute more to the next generation. The second term captures the transmission of within-group changes, weighted by group fitness. In expressing MLS using the Price equation in the form of Equation~\ref{eq:MLPrice} we do so to avoid the interpretation of other, causal, models such as Hamilton's~\cite{hamilton1963evolution} because we want the most general form of evolution, without presumption of {\it causality} or interpretations of {\it altruism}. As Equation~\ref{eq:MLPrice} is based solely on covariances, no causal presumptions are needed. %For example Hamilton's aim was to find an explicitly causal explanation connecting genes and altruism~\cite{queller1992general} whereas the formulation of Price is indifferent to the variety of causal processes that may affect fitness. This suits our ecological interpretation of the fitness and evolution of AI as it frees the interpretation from genes and altruism to include any correlative factors such as higher order interactions and other unknown interactions that may cause synergistic benefits.

In an organizational context, this framework shows how changes in performance stem from the interplay between competition among organizational units (departments, teams, firms) and selection processes within those units. Higher-performing groups expand their influence in the organization, while internal dynamics within groups also contribute to overall organizational evolution.

\subsubsection{Extending to AI-Augmented Organizations}
For AI-augmented organizations, we must explicitly distinguish between human and AI agents, as they may face different selection pressures and operate under different constraints:
\begin{equation}
\begin{split}
\overline{w} \, \Delta\overline{\pi} &= 
\underbrace{\text{Cov}_g(w^g,\overline{\pi}^g)}_{\text{between-group}} 
+ \underbrace{\mathbb{E}_g\left[\text{Cov}_{h \in g}(w_h^g,\pi_h^g)\right]}_{\text{within-group (human)}} 
+ \underbrace{\mathbb{E}_g\left[\text{Cov}_{ai \in g}(w_{ai}^g,\pi_{ai}^g)\right]}_{\text{within-group (AI)}} \\
&\quad + \underbrace{\mathbb{E}_g\left[\text{Cov}_{(h, ai) \in g}(w_h^g, \pi_{ai}^g)\right] 
+ \mathbb{E}_g\left[\text{Cov}_{(h, ai) \in g}(w_{ai}^g, \pi_h^g)\right]}_{\text{cross-agent interaction terms}}
\end{split}
\label{eq:ExtendedMLPrice}
\end{equation}
Here we decompose the change in average performance into between-group selection, within-group selection for both human and AI agents, and a fourth cross-agent interaction term. This extended formulation explicitly represents the covariance between human and AI influence and performance in the interaction term, highlighting several key challenges in AI governance, including different selection pressures, distinct fitness landscapes, and novel human-AI interaction dynamics that require careful institutional design. There is an extended description and analysis of this model in Appendix~\ref{app:MLS-economics} %{\color{red} Eq. 2 needs explanation. Also need to cover the mesoscopic level of the groups, agents are defined earlier, but groups as modules in games, clusters of games, firms, and the economy need expanding.}

\subsection{Aoki's Firm as an Endogenous Computation}
Aoki conceptualizes the firm as a distributed information-processing structure that maps signals about market states into coordinated actions through a modular game structure. Each module (department, algorithm, or human–AI hybrid) implements a local best-response function. Global coherence emerges when the institutional architecture induces an equilibrium that realizes super-modular complementarities \cite{Aoki2010}. Here, super-modular games are those games in which ``strategic complementarities'' arise -- i.e. when one agent takes a higher action, the others agents want to do the same. These games exhibit reinforcing behaviors, making them particularly relevant for modeling organizational complementarities.

%Formally, a game is super-modular if for any two strategy profiles $s' \geq s$ and any player $i$, the utility difference from increasing one's strategy is increasing in the strategies of other players (the set of players other than $i$ are denoted $-i$): $u_i(s'_i, s'_{-i}) - u_i(s_i, s'_{-i}) \geq u_i(s'_i, s_{-i}) - u_i(s_i, s_{-i})$. These games are those that also induce bifurcations (Thom's {\it Catastrophes}~\cite{thom1977structural}) in boundedly rational agents~\cite{harris2023smooth}.

The conceptual framing of Aoki's work that we leverage is the role of games in the structural organisation of the firm: agents playing games are a computational process, and that game structures and their relationship to one another encodes institutional norms~\cite{aoki2007endogenizing}. In this framing of the firm's institutions, Aoki says {\it an institution [is] a game-form ... the neo-institutionalists
like~\cite{North1990,north2010understanding} identify institutions with formal rules such as constitutions, statutory laws, and contracts, as well as informal rules such as social norms.} An alternative interpretation of a firm's institutions Aoki offers is that: {\it An institution [is] an endogenous equilibrium outcome of the game.} In his work Aoki favours reconciling these two {\it exogenous} and {\it endogenous} interpretations of institutions.

In formal terms, Aoki models the firm as a collection of interrelated games $\{G_1, G_2, ..., G_n\}$, where each game $G_i$ represents interactions within a particular domain (e.g., labor relations, supplier contracts, etc.). The institutional architecture of the firm defines the linkages between these games and the information flows that connect them. Equilibrium in this system represents a coherent organizational state in which all of the agents' beliefs and actions are mutually consistent \cite{Aoki2001}.

The computational capacity of the firm emerges from the collective operation of these local functions, modulated by the institutional rules that structure their interactions. For AI-augmented firms, super-modularity requires that human and AI components complement rather than substitute for each other, creating complementarities between human and AI decision processes that leverage the strengths of each while compensating for their weaknesses \cite{Brynjolfsson2017}.

\subsection{Evolutionary Behavioral Learning in Institutional Computation}
Building on Aoki's computational conception of the firm, we can model individual agents (human or AI) as boundedly rational learners whose behavioral repertoire evolves through repeated interaction within an institutionally scaffolded environment. Each agent learns to perform a local task and coordinate with other agents in ways that are bounded through shared behavioral conventions shaped by the institutional framework.

From an evolutionary game perspective, organizational behavior is viewed as the result of a computational learning process that has been institutionalized. Institutions offer both the fitness landscape -- through incentives, norms, and sanctions -- and the interpretive framework that determines the appropriateness or alignment of behavior with the firm's objectives. This dual role parallels North et al.'s analysis of how social orders create cooperation through rent-distribution systems that align elite interests with collective stability~\cite{NorthWallis2009}. This perspective is also grounded in the extensive literature on evolutionary game theory \cite{weibull1997evolutionary}.

For AI systems, this learning process is crucial to understand. AI agents adapt their behavior through gradient-based learning on millisecond timescales, while human behavioral adaptation typically occurs over days or weeks, and institutional rules evolve over months or years. These different timescales create complex dynamics that must be carefully managed through appropriate institutional structures \cite{rahwan2019machine, Russell2019}.

\subsection{Ostrom's Design Principles as Alignment Operators}
Ostrom identified eight design principles that underpin stable common-pool resource institutions. These principles function as alignment operators that shape the covariance structure in the multi-level Price equation, attenuating destructive within-group competition while preserving adaptive experimentation~\cite{ostrom2008design,wilson2013generalizing} (See Appendix~\ref{app:Ostrom-AI} for an interpretation for AI):

\begin{enumerate}
    \item \textbf{Clearly defined boundaries}: Defining who has rights to withdraw resources and the boundaries of the resource itself
    \item \textbf{Congruence with local conditions}: Rules appropriately match local needs and conditions
    \item \textbf{Collective-choice arrangements}: Those affected by rules can participate in modifying them
    \item \textbf{Monitoring}: Those who monitor resource conditions and behavior are accountable to users
    \item \textbf{Graduated sanctions}: Violations are punished based on seriousness and context
    \item \textbf{Conflict-resolution mechanisms}: Rapid, low-cost means for resolving conflicts
    \item \textbf{Minimal recognition of rights}: External authorities respect the right of users to self-govern
    \item \textbf{Nested enterprises}: When CPRs are parts of larger systems, governance activities are organized in multiple nested layers
\end{enumerate}

From an evolutionary perspective, Ostrom's principles can be understood as mechanisms that modify selection pressures to favor cooperative behaviors that enhance group fitness. By establishing clear boundaries, ensuring congruence with local conditions, and implementing graduated sanctions, these alignment operators create conditions where individual and collective interests become more aligned \cite{wilson2013generalizing}.

\section{Worked Example: Graduated Sanctions in the Prisoner's Dilemma \label{sec:PD-Inst-eg}}

How do institutions enforce cooperative behavior at the agent-to-agent level of interaction when noncooperative behavior has short term benefits? Here we illustrate the effect of Ostrom's Principle 5---\emph{graduated sanctions}---using the canonical Prisoner's Dilemma (PD) where payoffs represent years of jail time so that lower numbers are better for the players. The matrix is:
\[
\Pi_{\text{jail}} =
\begin{bmatrix}
(1, 1) & (3, 0) \\
(0, 3) & (2, 2)
\end{bmatrix}
\]

\begin{itemize}
    \item \textbf{Mutual cooperation} (C,C): 1 year each
    \item \textbf{Temptation to defect} (D,C): 0 years for the defector, 3 years for the sucker
    \item \textbf{Mutual defection} (D,D): 2 years each
    \item \textbf{Sucker's payoff} (C,D): 3 years for cooperator, 0 for defector
\end{itemize}

This matrix satisfies the PD condition in jail terms: $T < R < P < S$, and in utility terms (i.e., $-1 \times$ jail time): $T > R > P > S$. While mutual cooperation is clearly the optimal {\it collective} outcome where total jail time served is lowest, there is always the incentive for one agent to defect in the hope that the other agent will play the `sucker' and cooperate, so the Nash equilibrium of mutual defection is the collectively worst outcome for both agents where they collectively spend 4 years in jail. Note that cooperation is strictly dominated: no matter what the other player does, it is always beneficial for an agent to defect. If this were the entire story, then there would be very little to be said, but Ostrom's arguments allow us to expand on this example and illustrate how institutions can play a central role in the enforcement of cooperative strategies through both past and future sanctions. We illustrate this process using an exogenous (in Aoki's terms) institutional constraint: In the following example two agents play the Prisoner's dilemma repeatedly in an institutional context that enforces cooperation. 

\paragraph{Graduated‐sanction rule.}
Each player $i$ has a defection counter $k_i\in\{0,1,2,\dots\}$.  
When a player defects, their jail time is increased by a penalty \(\lambda k_i\), where \( \lambda>0 \) is an institutional severity parameter.

\paragraph{Sanctioned payoff matrix.}
Let $k_1$ be Player 1's current defection count and $k_2$ that of Player 2.  
The modified jail-time matrix with sanctions is:

\[
\Pi_{\text{s}}(k_1,k_2)=
\begin{bmatrix}
(1,\,1) &
(3,\,0+\lambda k_2)\\[6pt]
(0+\lambda k_1,\,3) &
(2+\lambda k_1,\,2+\lambda k_2)
\end{bmatrix}.
\]

The \( \lambda k_i \) are ``alignment parameters'' that modulate the payoff matrices and represent the institutional structure being imposed on the interactions between individual agents. The key properties here are:

\begin{itemize}
\item Sanctions are applied at the individual agent level: only defector $i$'s entry in each outcome receives the extra~$\lambda k_i$.
\item Punishment escalates with repeated violations: larger $k_i$ $\Rightarrow$ more jail time for that agent.
\item When \(\lambda k_i\) exceeds one year, the defector's temptation payoff
drops below the reward for mutual cooperation (\(0+\lambda k_i>1\)),
undermining the incentive to defect.
\end{itemize}

Note that once $\lambda k_i > 1$, defection becomes a strictly dominated strategy for player $i$: regardless of the other player's action, cooperation always yields a lower jail time. This demonstrates how institutional memory can reverse the incentive structure of the original dilemma, transforming it into a stability-enforcing coordination game. Ryan \textit{et al.}~\cite{ryan2016social} describe this transformation in game-theoretic terms as a shift in the \textit{effective game}—the actual game agents are playing once all relevant contextual factors are taken into account. These include the canonical payoff matrix as well as what they term \textit{social niche modifiers}: background conditions such as reputational systems, sanctioning norms, and institutional memory, all of which reshape agents' strategic landscapes. This moves the game away from being a context free interaction to being a game played in a social context that fits the agents into their local social or organizational niches. 

In this sense agents, whether human or AI, can enhance collective performance by actively having their interactions shaped by the social environment in which interactions occur via social niche construction~\cite{laland2016introduction}. Agents modulate group dynamics, incentives, and connections, effectively redesigning the ``game'' that is being played. This enables interventions that promote cooperation, coordination, and shared cognitive capacity within groups.

This model offers a demonstration of symmetric, history-dependent penalties that operationalize Ostrom's Principle~5: \emph{graduated sanctions} by embedding memory into the game, progressively shifting individual incentives toward sustained cooperation while preserving the possibility of forgiveness for first-time mistakes (i.e., when $k_i$ can be depreciated over time, such as a `forgetting' mechanism). In principle, one could also include a forgetting mechanism, where sanctions decay exponentially over time—analogous to reinforcement learning with memory loss. This would allow institutions to balance deterrence with leniency, ensuring that occasional lapses do not permanently bias future interactions.

Institutional sanctions reshape agents' expectations about non-cooperative behavior costs, affecting not just responses to past sanctions but predictions of future penalties. Forward-looking agents—especially AI systems that optimize over future trajectories—will proactively avoid behaviors that accumulate penalties, even if initial violations are lightly sanctioned. This predictive adaptation shifts equilibria toward cooperation by modifying the expected return landscape, discouraging even sophisticated agents from attempting to game the system through strategic defection.

\section{A Multi-Level Game-Theoretic Framework}

To formalize the connections between selection pressures at different organizational levels, we link the extended Price equation (Equation~\ref{eq:ExtendedMLPrice}) with the evolutionary dynamics of institutional rules. This establishes how the success of specific rule configurations—understood as institutional design choices—depends on the performance of the agents they coordinate. Crucially, these performance metrics are shaped not just by individual strategies, but by interaction structures and complementarities between heterogeneous agents, particularly between humans and AI systems.

\subsection{Institutional Games and Strategy Dynamics}

An Aoki-corporation is composed of a set of games $\{G_1, G_2, \ldots G_N\}$ being played by groups of agents within the organisation and these groups collectively interact to produce the corporation's economic output. Within each group $g$, we stylise the strategic interactions using the repeated Prisoner's Dilemma of Section~\ref{sec:PD-Inst-eg}, where there we modelled a single institutional rule based on punishing agents for a lack of cooperation, here labelled $j$. Each possible rule configuration determines an augmented payoff matrix through a transformation function that implements an alignment operator:

\begin{equation}
\Pi_j = T_j(\Pi_0)
\label{eq:RuleTransformation}
\end{equation}
where $\Pi_0$ is the baseline PD payoff matrix, i.e. $\Pi_{\text{jail}}$ in Section~\ref{sec:PD-Inst-eg}, and $T_j$ is the rule-induced transformation that introduces monitoring, sanctioning, and reward mechanisms in line with Ostrom's design principles (e.g., graduated sanctions or conflict resolution mechanisms). In Section~\ref{sec:PD-Inst-eg} $\Pi_j = \Pi_s(k_1,k_2)$. This transformation explicitly encodes the alignment parameters: How institutional rules modify interaction payoffs, facilitating both theoretical analysis and future empirical implementation.

Here we note that $\Pi_s(k_1,k_2)$ is a top down rule imposed by some regulatory mechanism within the firm, it likely has an implementation cost (see Section~\ref{sec:Selection-of-rules}). For illustration we contrast this with two alternative, bottom up, institutional rules: {\it tit-for-tat} and {\it win-stay, lose-switch}~\cite{nowak1993strategy}. Both of these rules have been proven effective in repeated PD and both are local rules one agent can follow in response to another agent's previous choices in order to enforce the social norm of cooperative behaviour. These three strategies: 1 top-down and two bottom-up, are then subject to evolutionary selection pressure based upon which is more successful within a firm.

Here we only consider one game $G_j$, the PD game modified by rule set $j$, i.e. the sanctioned payoff matrix of Section~\ref{sec:PD-Inst-eg}, and let $s_i \in \{C, D\}$ represent agent $i$'s strategy (cooperate or defect). The equilibrium behavior of agents is defined in the sense of Aoki's institutional equilibrium~\cite{Aoki2010}, formally characterized by:

\begin{equation}
s_i^* \in \arg\max_{s_i \in S_i} \mathbb{E}_{s_{-i} \sim \mu_j}[u_i(s_i, s_{-i})]
\label{eq:AokiEquilibrium}
\end{equation}
where $\mu_j$ is the probability distribution over the strategies of other agents $s_{-i}$ that emerges under institutional rule $j$.  Unlike standard Nash equilibrium, Aoki's institutional equilibrium recognizes that agents' beliefs and strategies are shaped by shared institutional knowledge and expectations about how others will behave under specific rule structures \cite{Aoki2001, greif2006institutions}. Individual strategies are locally optimal given the cognitive expectations and rule-enforced constraints of the institution, where these expectations include beliefs about monitoring, anticipated penalties, and the strategies of other agents under the shared rule structure.

We denote the resulting equilibrium payoffs as $\pi_h^*$ and $\pi_{ai}^*$ for humans and AI agents, respectively. These payoffs reflect not only direct material outcomes but also the effects of institutional constraints on agent behavior. Importantly, these equilibrium outcomes correspond directly to the performance measures $\pi_h^g$ and $\pi_{ai}^g$ that appear in our extended Price equation (Equation~\ref{eq:ExtendedMLPrice}), establishing the link between institutional design and multi-level selection pressures.

\subsection{Selection among Institutional Rule Sets \label{sec:Selection-of-rules}}

Each rule configuration $j$ generates a specific version of the institutional game $G_j$, which determines the equilibrium strategies and resulting performance metrics $\pi^*_j = (\pi^*_{h,j}, \pi^*_{ai,j})$. Note that in the example of Section~\ref{sec:PD-Inst-eg} the equilibrium we described for game matrix $\Pi_s(k_1,k_2)$ is an institutional equilibrium under Aoki's definition. We assume a well-mixed population of firms or organizations in which institutional rule configurations proliferate according to relative performance, absent spatial or network structure. Within this population, groups are effectively defined by their institutional arrangements—that is, group $g$ operates under rule configuration $j$, making the group's institutional structure its defining characteristic. We define $r_j$ as the frequency of configuration $j$ in the population of organizations.

The evolutionary dynamics over rule frequencies follow the replicator equation \cite{Taylor1978, Hofbauer1998}:

\begin{equation}
\dot{r}_j = r_j \left( V_j - \overline{V} \right)
\label{eq:RuleReplicator}
\end{equation}

\noindent where $\overline{V} = \sum_k r_k V_k$ is the average institutional fitness across all configurations. The fitness of rule configuration $j$ is given by:

\begin{equation}
V_j = f_j(\pi^*_j) - c_j
\label{eq:InstitutionalFitness}
\end{equation}

\noindent The function $f_j$ maps the agent-level equilibrium outcomes under rule $j$ into organizational fitness: %, incorporating the same structure as our group fitness function:
\begin{equation}
f_j(\pi^*_j) = \alpha \overline{\pi^*_{h,j}} + \beta \overline{\pi^*_{ai,j}} + \gamma \, \text{Cov}_{(h, ai) \in g_j}(\pi^*_{h}, \pi^*_{ai})
\label{eq:FitnessMapping}
\end{equation}
\noindent Here the terms $\overline{\pi^*_{h,j}}$ and $\overline{\pi^*_{ai,j}}$ are the average trait values, at equilibrium, for all humans and AI respectively under institutional rule configuration $j$. This instantiates a feedback loop in our evolutionary system: the average equilibrium performance of agents ($\overline{\pi^*_{h,j}}$ and  $\overline{\pi^*_{ai,j}}$) and their complementarity (captured by the covariance term) determine the fitness of rule configurations ($V_j$), which in turn shapes the future evolution of institutional rules through replicator dynamics. These institutional rules then influence the selection pressures on agents as captured in Equation~\ref{eq:ExtendedMLPrice}.

%Note that $f_j(\pi^*_j)$ parallels Equation~\ref{eq:GroupFitness}, reflecting the average fitness of organizations governed by rule $j$. This symmetry allows multi-level selection to be expressed within a unified evolutionary framework.

The cost term $c_j$ reflects the complexity and enforcement overhead required to implement the rule set. Drawing on Williamson's transaction cost economics \cite{williamson1979transaction,williamson2008outsourcing}, these costs arise from the fundamental problems of economic organization: bounded rationality, information asymmetry, and potential opportunism. Williamson argues that governance structures (including rule systems such as those covered in this article) evolve specifically to economize on these transaction costs through comparative institutional adaptation. In the current framework, the costs of institutional rules encompass multiple dimensions that directly influence their evolutionary fitness \cite{Williamson1985, North1990}:

\begin{itemize}
\item \textbf{Monitoring costs}: Resources required to observe agent behavior and system performance, including both human oversight and automated monitoring systems for AI agents
\item \textbf{Enforcement costs}: Administrative expenses of implementing sanctions, rewards, and conflict resolution mechanisms
\item \textbf{Coordination costs}: Communication overhead and complexity management as institutional rules become more sophisticated
\item \textbf{Adaptation costs}: Expenses associated with updating and maintaining institutional rules as technology and organizational needs evolve
\end{itemize}

More sophisticated institutional arrangements—such as those involving multi-level oversight, adaptive sanctioning systems, or complex human-AI coordination protocols—may generate higher fitness through enhanced cooperation and complementarity. However, they also incur higher implementation and maintenance costs, creating fundamental trade-offs that shape the evolutionary viability of different institutional configurations \cite{Ostrom2005}.

This framework connects agent-level strategy equilibria, institutional rule design, and selection dynamics at the population level of organizations. Rule configurations that promote high levels of cooperation and coordination—particularly those that effectively harness complementarities between humans and AI—will tend to proliferate in the population, provided their implementation costs do not outweigh the fitness advantages they confer. Conversely, rule systems that fail to align hybrid human-AI teams or that impose excessive administrative overhead will be outcompeted by more efficient institutional arrangements, even if they perform well in simpler, single-agent-type environments.

Bottom-up rules like tit-for-tat~\cite{nowak1992tit,milinski1987tit} offer an alternative approach to enforcement with lower organizational implementation costs but their own distinct trade-offs. Unlike top-down sanctioning mechanisms that require centralized monitoring and enforcement infrastructure, tit-for-tat operates through decentralized peer interactions, where each agent simply responds to its partner's previous action. This decentralized mechanism functions as an emergent alignment operator, transforming interaction structures to promote cooperative outcomes without formal institutional oversight. In essence, tit-for-tat represents how alignment parameters can arise endogenously through evolutionary processes rather than through explicitly designed institutional rules. This strategy is not cost-free—the costs are instead borne locally by the agents themselves, who must accept occasional defections and retaliate accordingly, potentially sacrificing short-term gains. 

Moreover, these rules must diffuse through the population via social learning or be discovered by random chance rather than being immediately and universally adopted. However, this diffusion process enables parallel exploration of multiple rule sets simultaneously, creating evolutionary competition between different coordination strategies. In the long run, tit-for-tat's effectiveness comes from its competitive average payoff and its simplicity—agents need only remember one previous interaction rather than maintain complex institutional memory. This balance between immediate costs and long-term benefits exemplifies how evolutionary processes can select for cooperation-enforcing mechanisms even without centralized authority \cite{nowak1993strategy, Axelrod1984}.

Taken together, this framework embeds multi-level selection within a tractable evolutionary game-theoretic structure, bridging the Price equation at the micro level of individual agents and replicator dynamics at the meso level of institutional evolution. Institutional rules serve as the crucial link between strategic behavior and organizational evolution, functioning as alignment operators that shape the selection pressures captured in our extended Price equation while themselves being subject to selection based on the organizational outcomes they generate.

\section{Empirical Case Studies}

To base our theoretical framework in real-world contexts, we consider two case studies of organizations that have integrated AI systems under different institutional arrangements:

\subsection{Algorithmic Trading Firms}
Algorithmic trading represents one of the earliest domains where AI agents became key decision-makers within economic organizations. We looked at how trading firms have evolved institutional rules to manage the risks of high-frequency trading algorithms while preserving their speed advantages. Some of the key elements of modern algorithmic trading include:
\begin{itemize}
    \item Circuit breakers and position limits function as Ostrom-style graduated sanctions, automatically constraining algorithmic behavior when predefined risk thresholds are exceeded~\cite{kim2004makes}
    \item Firms with transparent internal oversight processes show lower incidence of ``flash crash'' events, consistent with the monitoring principle: {\it This Note proposes instead that the largest high-frequency traders be subject to heightened regulatory oversight to ensure fair dealing.}~\cite{barrales2012lessons}
    \item Rule evolution accelerated following major market disruptions, demonstrating punctuated equilibrium dynamics in institutional adaptation~\cite{kirilenko2013moore,true2019punctuated}
\end{itemize}

These findings illustrate how financial institutions have developed nested oversight mechanisms that operate at multiple timescales—from millisecond-level circuit breakers to daily risk reviews to quarterly governance updates~\cite{Kirilenko2017}. The evolution of trading firm governance demonstrates the importance of Ostrom's eighth principle—nested enterprises. Effective governance operates at multiple levels: internal firm risk controls, exchange-level circuit breakers, and regulatory frameworks. This nested structure ensures that failures at one level can be contained before they cascade through the entire system \cite{Budish2015}.

\subsection{Amazon's Automated Scheduling Failure}
A specific example of rule exploitation occurred when the scheduling algorithm began assigning workers to 'back-to-back' shifts with less than 8 hours between them, technically complying with corporate policies while maximizing certain efficiency metrics. This practice led to increased absenteeism, safety incidents, and eventually worker protests, illustrating how AI systems can identify and exploit loopholes that human managers typically would not \cite{Lecher2019, wood2021algorithmic}. Some key observations:
\begin{itemize}
    \item Absence of collective-choice arrangements led to worker pushback, as those affected by the algorithm had no input into its design or operation
    \item Lack of congruence with local conditions resulted in impractical schedules that failed to account for transportation constraints and family obligations
    \item Missing conflict-resolution mechanisms prevented timely adjustments to problematic schedules
    \item The system was eventually abandoned, demonstrating negative selection against rule systems that violate Ostrom principles
\end{itemize}

This case illustrates how even technologically sophisticated organizations face institutional selective pressures when AI deployment conflicts with social dynamics and human needs. Amazon's algorithm optimized for operational efficiency but failed to incorporate local knowledge and stakeholder input, ultimately making it unsustainable \cite{Kantor2021}.

% Add these two subsections to Section 5 Empirical Case Studies

\subsection{ICANN: Global Multistakeholder Technology Governance}

The Internet Corporation for Assigned Names and Numbers (ICANN) demonstrates successful implementation of polycentric governance for global technology systems. Since 1998, ICANN has coordinated the Internet's domain name system through a multistakeholder model involving governments, private sector, civil society, and technical experts~\cite{scholte2021legitimacy}.

ICANN's structure implements several Ostrom principles directly:

\begin{itemize}
\item \textbf{Clearly defined boundaries} through formal stakeholder constituencies with specific roles and authority
\item \textbf{Collective choice arrangements} via policy development processes where affected parties participate in rule-making
\item \textbf{Monitoring mechanisms} through transparent meetings and regular independent reviews
\item \textbf{Graduated sanctions} from community pressure to formal accountability mechanisms, including board removal powers
\end{itemize}

The 2016 IANA stewardship transition illustrates institutional evolution under selection pressure. When the U.S. government withdrew from oversight functions, ICANN's community developed enhanced accountability mechanisms while preserving core governance principles~\cite{icann2016transition}. This adaptive response demonstrates how institutional rules can evolve to maintain legitimacy as external conditions change.

\textbf{Nested governance structures} enable coordination across multiple levels: individual constituencies, supporting organizations, the Board, and external relationships with governments and technical communities. This polycentric arrangement allows local autonomy while maintaining global coordination---exactly the balance Ostrom identified as crucial for commons governance.

\subsection{Maine Lobster Fisheries: Bottom-Up Resource Management}

Maine's lobster fishery exemplifies self-organizing governance that emerged without formal institutional design. Fishing communities developed territorial systems through repeated interactions, creating effective resource management that combines local knowledge with state-level coordination~\cite{wilson2007precursors,acheson2003capturing}.

\textbf{Harbor gangs} establish informal territorial boundaries defended through graduated sanctions: verbal warnings, trap interference, gear destruction, and ultimately exclusion from fishing areas. These territories emerge from lobster biology (sedentary behavior) and fishing technology (trap-based), creating conditions favoring local stewardship over mobile extraction.

The system demonstrates multiple Ostrom principles:

\begin{itemize}
\item \textbf{Boundary definition} through territorial arrangements aligned with ecological and social conditions
\item \textbf{Monitoring} via dense social networks that share information about resource conditions and rule violations
\item \textbf{Conflict resolution} combining informal mediation with formal state procedures
\item \textbf{Nested enterprises} integrating local territorial rules with state advisory councils and regulations
\end{itemize}

From a multi-level selection perspective, harbor gangs with effective cooperation and conservation achieve higher long-term fishing success, creating selection pressure for institutional arrangements that balance individual success with collective stewardship~\cite{brewer2012boundaries}. The system has adapted over time to changing lobster abundance and climate pressures while maintaining core governance principles.

Both cases validate that Ostrom-style governance can operate effectively from local resource management to global technology coordination, providing empirical support for applying these principles to AI institutional design.

\section{Inclusive Institutions and AI Governance}

This section explores how Acemoglu's institutional economics framework applies to AI governance within firms, focusing on the distinction between inclusive and extractive institutional arrangements.

\subsection{Inclusive vs. Extractive AI Institutions}

Acemoglu's distinction between inclusive and extractive institutions offers a valuable lens for analyzing AI governance systems. While inclusive institutions distribute economic and political power broadly across society, extractive ones concentrate power among elites. The framework below characterizes institutions along multiple dimensions of inclusivity versus extractiveness, helping us understand how institutional structures shape AI deployment and outcomes.

\begin{table}[h]
\centering
\begin{tabular}{@{}lll@{}}
\toprule
\textbf{Dimension} & \textbf{Inclusive AI Institutions} & \textbf{Extractive AI Institutions} \\
\midrule
Transparency & High algorithmic transparency & Black-box decision-making \\
Participation & Stakeholder input in design & Centralized control \\
Value distribution & Broadly shared benefits & Concentrated gains \\
Adaptation rights & Open modification & Closed proprietary systems \\
Data governance & User control & Corporate/state extraction \\
Oversight & Distributed accountability & Centralized or absent oversight \\
Knowledge access & Broad algorithmic literacy & Technical gatekeeping \\
\bottomrule
\end{tabular}
\caption{Characteristics of inclusive versus extractive AI institutions}
\label{tab:inclusive_extractive}
\end{table}

\paragraph{Polycentric governance and institutional adaptability.}
Inclusive AI institutions rarely emerge from a single centre of authority.
Instead, they evolve within a \emph{polycentric} ecosystem in which local, regional, and sector-specific bodies enact partially overlapping rules and monitoring systems. Such arrangements distribute veto power, encourage contextual experimentation, and create evolutionary competition among rule configurations, thereby mitigating institutional lock-in and elite capture risks\citep{OstromPolycentric2010}. Embedding AI governance within a
polycentric architecture therefore strengthens the between-group selection
mechanisms identified in Section \ref{sec:mls}, aligning technological
innovation with broad social welfare rather than narrow rent extraction.

Acemoglu's framework suggests that inclusive AI institutions would foster greater innovation and more sustainable economic growth by distributing the benefits of AI more broadly \cite{Acemoglu2012, Acemoglu2021harms}. However, the tendency toward increasing returns to scale in AI development may naturally push toward more extractive arrangements unless countered by deliberate institutional design \cite{Korinek2021}.

The distinction between inclusive and extractive AI institutions has important implications for our multi-level selection framework. Extractive institutions tend to favor within-group selection, allowing powerful actors to capture disproportionate benefits. In contrast, inclusive institutions strengthen between-group selection by aligning individual incentives with collective outcomes \cite{Acemoglu2012, Bowles2004}.

\subsection{Firms as Institutional Economies}

Firms, like nations, constitute economies of rules—this represents an isomorphism between nested institutional forms at micro (firm) and macro (state) levels. Acemoglu points out that ``some firms operate under fairly inclusive principles, [while] others are much more in the mode of extractive institutions''~\cite{Acemoglu2012}. This isomorphism justifies importing concepts like extractive/inclusive institutions into corporate contexts: corporate policies governing profit sharing, promotion, data rights, and algorithmic control all define institutional structures that affect internal fitness.

Within our multi-level selection framework, firms can be understood as organizational units subject to selection pressures at multiple levels. The internal institutional architecture of a firm shapes the incentives of individual agents (human and algorithmic) and mediates the covariance structure between individual and organizational fitness. Inclusive institutions within firms enhance between-group selection by aligning individual and collective interests, while extractive institutions prioritize within-group selection, potentially sacrificing long-term adaptation for short-term gains by powerful actors \cite{henrich2015secret, Wilson2014}.

As AI systems become embedded in corporate decision-making, the internal institutional architecture of firms plays an increasing role in shaping how AI is deployed, governed, and evolved. Inclusive firms are more likely to implement AI systems that augment human capabilities, respect stakeholder values, and distribute benefits more evenly. They are also more likely to anticipate and mitigate strategic failures arising from the misalignment of AI incentives \cite{acemoglu2023power, rahwan2019machine}.

Extractive firms, by contrast, may pursue AI adoption strategies that prioritize cost-cutting or surveillance, often without institutional checks. This can lead to brittle performance, backlash, and reputational risk \cite{zuboff2019age}. In evolutionary terms, inclusive institutions serve as alignment operators at the firm level: they regulate the incentive gradients on which AI systems are trained, embedding long-run human values into near-term optimization.

The path-dependent nature of institutional development has significant implications for AI governance. Early decisions about AI deployment create lock-in effects through technical standards, data advantages, expertise concentration, and regulatory precedents. Once established, these patterns become increasingly difficult to modify, highlighting the importance of thoughtful institutional design at early stages of AI integration \cite{North1990, Arthur1994}.

\section{Practical Policy Recommendations and Further Work}
\label{sec:implementation}

While the preceding sections develop a unified evolutionary–institutional theory of AI governance, they have intentionally left the specification of concrete policy levers open.
Here we outline an initial agenda that demonstrates \emph{how} the theoretical
principles in Sections \ref{sec:Theo-Found}–\ref{sec:PD-Inst-eg} can be operationalised
without sacrificing their generality.

\subsection{Top--Down Redirection Levers}
\label{subsec:topdown}

Acemoglu's programme~\citep[][pg. 23]{AcemogluRedesignAI2021} rests on three mutually reinforcing
pillars: \emph{Government involvement, norm shifting, and oversight}, that can steer artificial-intelligence development away from
``so-so automation'' and toward broad, human-centred prosperity%
~\citep{AcemogluRedesignAI2021}:
\begin{enumerate}[label=(\alph*)]
  \item \textbf{Correcting distorted market incentives.}  Equalise the
        effective tax burden on labour and capital, and direct public
        R\&D subsidies and tax credits toward \emph{human-complementary} AI
        so that firms face no automatic cost advantage in pursuing
        displacement-only automation.
  \item \textbf{Redirecting the research agenda.}  Reform professional norms,
        funding criteria and university curricula so that prestige and
        resources flow to projects that create new tasks and skills for
        workers—e.g.\ through interdisciplinary ``human-centred AI’’
        labs—rather than narrowly optimising prediction benchmarks.
  \item \textbf{Reinforcing democratic oversight and worker power.}  Strengthen
        collective-bargaining rights, mandate algorithmic transparency and
        independent audits, and employ antitrust and labour-market regulation
        to curb excessive concentration and surveillance abuses.
\end{enumerate}

Because these levers operate at nation-state scale, they confront the
\emph{political constraint} that otherwise locks societies into
path-dependent, extractive trajectories~\citep[][ch.~13]{AcemogluRobinson2012}.
``Procurement -- based scoring'' systems (e.g. explicitly weighting bids by their projected contribution to human–AI complementarity) and new ``algorithmic liability'' rules (e.g. holding corporations deploying AI systems are held strictly or proportionally liable for harms caused) have been noted as \emph{possible tools}, but they are not part of the three core pillars articulated by Acemoglu.

\subsection{Bottom--Up Capability Levers}
\label{subsec:bottomup}

Complementing top-down incentives, Ostromian commons governance places
implementation responsibility in the hands of those most affected by the rules.
Recent pilot work on \emph{Responsible LLM Use in Higher Education} demonstrates
how a workshop curriculum can instantiate all eight of Ostrom’s design
principles in a research-intensive environment\citep{Harre2025Responsible}.
We note that (a) above ``human--complementary AI'', (b) above ``projects that create new tasks
and skills for workers'', and (c) above ``worker power'' are central to this bottom up educational program and align with Ostrom's design principles. They are also intended to complement and mutually support centralized policies in a hierarchical fashion. Other key elements include:
\begin{itemize}
  \item \emph{Boundary setting.} Participants articulate phase-specific guard-rails
        on acceptable AI assistance (Principle 1);
  \item \emph{Peer monitoring.} Reciprocal verification of citations and data
        provenance reduces policing cost while increasing collective trust
        (Principle 4);
  \item \emph{Graduated sanctions.} The curriculum’s ``response protocols'' scale
        interventions from mild correction to mandatory disclosure, mirroring
        the graduated-sanction model developed in Section~\ref{sec:PD-Inst-eg}.
\end{itemize}

\subsection{Polycentric Experimentation Agenda}
\label{subsec:polycentricagenda}

A polycentric configuration of governance centres~\citep{ostrom1980artisanship,OstromPolycentric2010,ostrom2004multi}
offers two strategic advantages:
\begin{enumerate}[label=(\roman*)]
  \item \textbf{Adaptive fit.} Multiple partially overlapping jurisdictions
  experiment with context-specific rule-sets, shortening the discovery time for locally effective alignment operators;
  \item \textbf{Resistance to capture.}     Dispersed veto points raise the cost of centralised manipulation, maintaining the inclusive institutional properties mapped in Table~\ref{tab:inclusive_extractive}.
\end{enumerate}
Future work should establish standardised \emph{metrics of institutional fitness}
(e.g.\ complementarity indices, cross-term covariances from Equation \eqref{eq:ExtendedMLPrice})
and compare them across experimental sites.

\subsection{Open Research Questions}

\begin{enumerate}[label=\arabic*.]
  \item \textbf{Measurement.} How can the cross-agent covariance terms developed
        in Appendix B be operationalised with real organisational data?
  \item \textbf{Scaling laws.} Do the benefits of inclusive AI institutions
        exhibit diminishing or increasing returns as firm size grows?
  \item \textbf{Translational infrastructure.} What governance APIs and data
        standards are required for firms to plug modular AI components into
        existing incentive architectures without re-engineering entire rule
        stacks?
\end{enumerate}

Further work that can address these questions will turn the theoretical framework presented here into a practical and adaptive programme of policy experiments, closing the ``implementation gap'' central to policy discussions while preserving the evolutionary logic that motivates this work.

% ==============================================================
%  INSERT THE FOLLOWING PARAGRAPH *IMMEDIATELY AFTER* TABLE 2
%  IN SECTION 6 “Inclusive vs Extractive AI Institutions”
% ==============================================================

\section{Discussion}

Multi-level selection theory provides a powerful framework for understanding the evolution of institutions that govern AI systems within economic organizations. By recognizing that selection operates simultaneously at the levels of individual agents, teams, organizations, and institutional rules, we can better understand the complex dynamics that shape AI deployment and governance.

Aoki's computational theory of the firm provides the micro-foundational architecture through which these selection pressures operate, while Ostrom's design principles offer empirically grounded alignment operators that can stabilize cooperation even in competitive environments. Their integration yields a framework for adaptive safety—institutional structures that keep AI systems aligned with human values even as they learn and environments change.

The incorporation of Acemoglu's insights on inclusive versus extractive institutions further develops this framework by highlighting how power dynamics shape institutional evolution. Achieving robust AI alignment requires attention not only to technical design but also to the distribution of economic and political power that determines which institutional arrangements prevail. By situating AI alignment within this broader context of institutional evolution and multi-level selection, we offer a holistic approach to AI governance that works to develop the technical, social, and economic dimensions simultaneously. 

Such an integrated perspective extends beyond just technical fixes, acknowledging that effective AI governance is born from the complex interaction of formal rules, informal norms, technical systems, and power dynamics within economic organizations. Furthermore, managerial strategies that involve altering organizational structures and setting incentives for individuals and teams are essentially experimental acts to enhance a firm's overall performance. 

In that sense the underlying principles of this paper are a formal description of the process of development and improvement that any corporation undertakes in order to survive in a competitive environment, adjusted for the specific issues related to AI and its nascent artificial agency. Similarly, these policy recommendations are not very different to those that have been implemented previously and that are needed to avoid localized competition and optimization from causing negative social externalities.

The case studies we have examined demonstrate that these theoretical principles have practical relevance across diverse organizational contexts. From algorithmic trading firms to healthcare organizations to Amazon's scheduling system, Ostrom-style governance principles help create more robust, adaptive, and inclusive AI-augmented institutions.

As AI systems become increasingly autonomous and central to economic decision-making, developing appropriate institutional structures for their governance becomes ever more critical. The framework we have presented offers a theoretically grounded yet practically oriented approach to this challenge—one that recognizes the fundamental evolutionary dynamics that shape institutional outcomes while providing concrete design principles for practitioners.

\appendix

\section*{Appendix}

\section{Design Principles for AI-Compatible Institutions \label{app:Ostrom-AI}}

This article has argued for merging Ostrom's principles with AI design guidelines in order to strengthen AI governance. This approach will foster cooperation and alignment at each organizational level. A proposed list of AI counterparts to Ostrom's principles is provided, though these should undergo an adaptive, evolutionary, process of their own as agentic AI progresses:

\begin{enumerate}
    \item \textbf{Clearly defined boundaries}: Explicit delineation of AI and human responsibilities and authority, with transparent documentation of system capabilities and limitations
    
    \item \textbf{Congruence with local conditions}: AI systems designed to complement local knowledge and practices, with adaptation mechanisms to accommodate contextual variation
    
    \item \textbf{Collective-choice arrangements}: Stakeholder participation in AI design and governance decisions, including those who will be affected by algorithmic outputs
    
    \item \textbf{Monitoring}: Transparent and accessible monitoring of AI behavior and impacts, with attention to both intended outcomes and unintended consequences
    
    \item \textbf{Graduated sanctions}: Escalating interventions for AI systems that violate boundaries, from mild corrections to complete deactivation based on violation severity
    
    \item \textbf{Conflict-resolution mechanisms}: Efficient processes for addressing disputes over AI decisions, including human override capabilities.
    
    \item \textbf{Minimal recognition of rights}: External authorities respect the autonomy of local AI governance while providing supportive legal and regulatory frameworks
    
    \item \textbf{Nested enterprises}: Multi-level governance spanning from local deployment to industry standards, with appropriate coordination across levels
    
    \item \textbf{Algorithmic transparency}: AI systems designed to be interpretable to relevant stakeholders, with appropriate documentation of training data, objective functions, and decision logic
    
    \item \textbf{Value alignment mechanisms}: Explicit processes to ensure AI optimizes for institution-wide objectives rather than narrow metrics, including oversight of objective function design
    
    \item \textbf{Adaptive experimentation}: Controlled testing of AI capabilities with systematic learning from errors, allowing for innovation while containing potential harms
\end{enumerate}

These are the alignment operators discussed in thew main body of this article and will aid in the creation of institutional structures that govern increasingly autonomous AI systems while preserving adaptability and performance. In the same way that Ostrom's principles enable sustainable management of common-pool resources through institutional arrangements that align individual and collective interests, these AI-compatible design principles promote alignment between algorithmic optimization and broader organizational goals. They function by shaping selection pressures at both individual and organizational levels, promoting evolutionary dynamics that favor cooperative human-AI interactions \cite{rahwan2019machine, Russell2019, Ostrom2005}.

This approach bridges several important perspectives in AI governance. First, it extends Rahwan et al.'s {\it machine behavior} framework, which advocates studying AI systems as autonomous agents with emergent behaviors requiring multi-disciplinary analysis. Where Rahwan provides a mesoscopic view (between microscopic neural networks and macroscopic economic outputs) for understanding AI behavior, these institutional alignment operators offer an evolutionary mechanism for actively shaping agent-to-agent interactions through selection pressures. Second, it provides concrete institutional implementations of Russell's value alignment principles. While Russell emphasizes designing individual AI systems with inherent uncertainty about human objectives, our framework scales this point to the organizational level via institutional structures.

This evolutionary approach extends both of these perspectives by introducing selection dynamics absent from their analyses. Rather than focusing solely on engineering better-aligned individual systems (Russell) or observing their emergent behaviors (Rahwan), we demonstrate how institutional rules create evolutionary pressures that naturally select for aligned systems through competitive dynamics. The replicator equation governing rule adoption (Equation~\ref{eq:RuleReplicator}) formalizes how alignment-promoting institutional designs proliferate when they generate higher organizational fitness, while misaligned arrangements are selected against. This evolutionary mechanism provides a complementary perspective to Russell's direct design approach and Rahwan et al's behavioral analysis, as it exploits market and organizational selection to continuously improve alignment mechanisms over time.

By explicitly modeling human-AI interactions through cross-agent covariance terms (Equation~\ref{eq:CovarianceDefinition}), we address Rahwan et al's call for approaches that capture complex interdependencies between human and machine behaviors. Our multi-level selection framework formalizes these interactions, showing that the complementarity between humans and AI itself can be made subject to selection pressures that favor aligned collaboration. This builds an evolutionary pathway to human-AI {\it complementarity} rather than {\it substitution}, addressing one of Russell's central concerns about AI development trajectories.

\section{Technical Analysis of Multi-Level Selection Terms \label{app:MLS-economics}}
\label{app:selection-terms}

This appendix provides detailed technical analysis of the selection terms in our extended Price equation (Equation~\ref{eq:ExtendedMLPrice}). We examine the first three terms that capture within-type selection dynamics, followed by the cross-agent interaction terms that represent the novel aspects of human-AI organizational evolution.

\subsection{Conceptual Distinction: Group Fitness vs. Group Performance}

To properly interpret the multi-level selection framework, we must clarify the conceptual mapping from evolutionary biology to organizational economics:

\begin{itemize}
\item \textbf{Group Fitness} ($w^g$): Analogous to reproductive success in biology, this represents the group's ability to persist, grow, and expand its influence within the organizational population. Fitness captures whether the group's institutional practices and characteristics will be represented in future organizational states through mechanisms such as budget allocation, personnel expansion, replication of practices, or acquisition of resources \cite{Nelson1982}.

\item \textbf{Group Performance} ($\pi^g$): Analogous to phenotypic traits in biology, this represents the observable characteristics and outputs of the group—productivity, quality, coordination effectiveness, or goal achievement. Performance measures how well the group executes its assigned functions within the organization \cite{hackman2002leading}.
\end{itemize}

\noindent The central insight from multi-level selection theory is that performance (like biological traits) and fitness (like reproductive success) are distinct and potentially divergent \cite{Frank1998, Wilson2007}. A high-performing group may have low fitness if organizational structures prevent its practices from spreading, while a mediocre-performing group might achieve high fitness through institutional advantages. The covariance between performance and fitness—captured in the between-group selection term—determines whether organizational evolution favors groups that perform well at their tasks or groups that are simply effective at securing resources and expanding influence.

\subsection{Within-Type Selection Terms: Technical Definitions}

The first three terms in our extended Price equation represent classical multi-level selection dynamics adapted for human-AI organizations:

\begin{enumerate}
\item \textbf{Between-group selection}: $\text{Cov}_g(w^g,\overline{\pi}^g)$
\item \textbf{Within-group selection (human)}: $\mathbb{E}_g\left[\text{Cov}_{h \in g}(w_h^g,\pi_h^g)\right]$
\item \textbf{Within-group selection (AI)}: $\mathbb{E}_g\left[\text{Cov}_{ai \in g}(w_{ai}^g,\pi_{ai}^g)\right]$
\end{enumerate}

\subsection{Between-Group Selection Analysis}
\subsubsection{Technical Definition}
The between-group selection term $\text{Cov}g(w^g,\overline{\pi}^g)$ measures the covariance between group fitness ($w^g$) and average group performance ($\overline{\pi}^g$) across all groups in the population. The covariance is calculated as:

\begin{equation}
\text{Cov}_g(w^g,\overline{\pi}^g) = \frac{1}{n_g}\sum_{g=1}^{n_g}(w^g - \mathbb{E}_g[w^g])(\overline{\pi}^g - \mathbb{E}_g[\overline{\pi}^g])
\label{eq:BetweenGroupCov}
\end{equation}

\noindent where $\overline{w} = \frac{1}{n_g}\sum_{g=1}^{n_g} w^g$ and $\overline{\pi} = \frac{1}{n_g}\sum_{g=1}^{n_g} \overline{\pi}^g$ represent the population means of group fitness and group performance, respectively. Here, $n_g$ is the number of groups, human and AI, in the population.
\subsubsection{Interpretation}
\paragraph{Positive Covariance}
Indicates that groups with higher average performance tend to have higher fitness (growth rates, resource acquisition, expansion). This creates group-level selection pressure favoring:
\begin{itemize}
\item Organizational units (departments, divisions, firms) with superior collective performance
\item Institutional arrangements that promote coordination and cooperation \cite{Ostrom2005}
\item Management practices that enhance group-level outcomes \cite{hackman2002leading}
\end{itemize}
\paragraph{Negative Covariance}
Suggests that higher-performing groups actually have lower fitness, possibly indicating:
\begin{itemize}
\item Simpson's paradox effects where individual excellence undermines group cohesion \cite{Simpson1951}
\item Resource constraints that penalize high-performing groups through taxation or redistribution
\item Institutional arrangements designed to equalize outcomes across groups
\end{itemize}
\paragraph{Zero Covariance}
Implies no systematic relationship between group performance and group fitness, suggesting:
\begin{itemize}
\item Random or externally determined group success
\item Perfect compensation mechanisms that equalize group fitness regardless of performance
\item Measurement artifacts where performance and fitness capture orthogonal organizational aspects
\end{itemize}

\subsubsection{Practical Examples}
Common manifestations of between-group selection include:
\begin{itemize}
\item \textbf{Corporate divisions}: High-revenue divisions receive increased investment and operational autonomy
\item \textbf{Academic departments}: Well-performing departments obtain larger budgets and faculty hiring allocations
\item \textbf{Military units}: Successful units receive superior equipment and strategic assignments
\item \textbf{Research teams}: Productive teams gain access to better resources and collaboration opportunities
\end{itemize}

\subsection{Within-Group Selection: Human Agents}

\subsubsection{Technical Definition}
The term $\mathbb{E}_g\left[\text{Cov}_{h \in g}(w_h^g,\pi_h^g)\right]$ measures the expected covariance between human fitness and human performance within groups, averaged across all groups in the population.

\subsubsection{Computational Form}
\begin{equation}
\mathbb{E}_g\left[\text{Cov}_{h \in g}(w_h^g,\pi_h^g)\right] = \frac{1}{n_g}\sum_{g=1}^{n_g} \text{Cov}_{h \in g}(w_h^g,\pi_h^g)
\label{eq:HumanWithinGroup}
\end{equation}

For each group $g$, the within-group covariance is:
\begin{equation}
\text{Cov}_{h \in g}(w_h^g,\pi_h^g) = \frac{1}{|H_g|}\sum_{h \in H_g} (w_h^g - \overline{w}_h^g)(\pi_h^g - \overline{\pi}_h^g)
\label{eq:HumanGroupCovariance}
\end{equation}

where $H_g$ denotes the set of human agents in group $g$, and $\overline{w}_h^g$, $\overline{\pi}_h^g$ represent the group-specific means for humans.

\subsubsection{Interpretation}

\paragraph{Positive Covariance}
Indicates that higher-performing humans tend to have higher fitness within groups, manifesting through:
\begin{itemize}
\item \textbf{Meritocratic promotion}: Superior performers advance faster in organizational hierarchies
\item \textbf{Resource access}: High performers receive better tools, training, and development opportunities
\item \textbf{Influence accumulation}: Successful individuals gain increased decision-making authority and autonomy
\item \textbf{Reputation effects}: Performance translates into professional recognition and career advancement \cite{sauder2012status}
\end{itemize}

\paragraph{Negative Covariance}
Suggests that better-performing humans have lower fitness, potentially due to:
\begin{itemize}
\item \textbf{Tall poppy syndrome}: Organizations penalize standout performers to maintain group harmony \cite{Feather1989}
\item \textbf{Resource redistribution}: High performers are "taxed" to support struggling colleagues
\item \textbf{Risk assignment}: Top performers receive more challenging and potentially career-damaging assignments
\item \textbf{Exploitation dynamics}: Capable individuals are overworked without proportional rewards
\end{itemize}

\subsubsection{Measurement Considerations}
\paragraph{Fitness Proxies for Humans}
Human fitness ($w_h^g$) can be operationalized through:
\begin{itemize}
\item Promotion rates and career advancement velocity
\item Salary increases and bonus allocations
\item Peer influence measures and network centrality \cite{Burt2005}
\item Discretionary resource allocation and budget authority
\item Performance evaluation scores and peer ratings
\end{itemize}

\paragraph{Performance Metrics for Humans}
Human performance ($\pi_h^g$) encompasses:
\begin{itemize}
\item Productivity measures (output quantity and quality)
\item Innovation indices (patents, ideas generated, process improvements)
\item Customer satisfaction ratings and client retention
\item Collaboration effectiveness and team contribution scores
\item Goal achievement and project completion rates
\end{itemize}

\subsection{Within-Group Selection: AI Agents}

\subsubsection{Technical Definition}
The term $\mathbb{E}_g\left[\text{Cov}_{ai \in g}(w_{ai}^g,\pi_{ai}^g)\right]$ measures the expected covariance between AI fitness and AI performance within groups, averaged across all groups.

\subsubsection{Computational Form}
\begin{equation}
\mathbb{E}_g\left[\text{Cov}_{ai \in g}(w_{ai}^g,\pi_{ai}^g)\right] = \frac{1}{n_g}\sum_{g=1}^{n_g} \text{Cov}_{ai \in g}(w_{ai}^g,\pi_{ai}^g)
\label{eq:AIWithinGroup}
\end{equation}

For each group $g$:
\begin{equation}
\text{Cov}_{ai \in g}(w_{ai}^g,\pi_{ai}^g) = \frac{1}{|A_g|}\sum_{ai \in A_g} (w_{ai}^g - \overline{w}_{ai}^g)(\pi_{ai}^g - \overline{\pi}_{ai}^g)
\label{eq:AIGroupCovariance}
\end{equation}

where $A_g$ denotes the set of AI agents in group $g$, and $\overline{w}_{ai}^g$, $\overline{\pi}_{ai}^g$ represent the group-specific means for AI agents.

\subsubsection{Interpretation}

\paragraph{Positive Covariance}
Indicates that better-performing AI systems tend to have higher fitness, manifesting through:
\begin{itemize}
\item \textbf{Deployment frequency}: Superior AI systems are utilized more frequently across organizational tasks
\item \textbf{Resource allocation}: High-performing AI receives increased computational resources, memory, or processing priority
\item \textbf{Update priority}: Successful AI systems receive more frequent improvements and optimizations
\item \textbf{Expansion scope}: Well-performing AI tools are deployed to additional use cases and user groups
\item \textbf{Investment focus}: Organizations allocate more R\&D resources to enhancing already-successful AI systems \cite{Brynjolfsson2017}
\end{itemize}

\paragraph{Negative Covariance}
Suggests that better-performing AI systems have lower fitness, possibly due to:
\begin{itemize}
\item \textbf{Replacement pressure}: Highly successful AI may trigger concerns about over-dependence, leading to diversification efforts
\item \textbf{Resource saturation}: Top-performing AI may already be optimally resourced, with marginal improvements yielding diminishing returns
\item \textbf{Integration challenges}: Highly capable AI may be difficult to integrate with existing human workflows
\item \textbf{Risk management}: Organizations may deliberately limit deployment of powerful AI systems due to safety or control concerns \cite{Russell2019}
\end{itemize}

\subsubsection{Distinctive Aspects of AI Selection}
AI agent selection differs fundamentally from human selection in several ways:

\paragraph{Temporal Dynamics}
\begin{itemize}
\item \textbf{Rapid evolution}: AI systems can be updated and modified orders of magnitude faster than human capabilities evolve
\item \textbf{Instant replication}: Successful AI algorithms can be immediately copied across multiple organizational contexts
\item \textbf{Continuous adaptation}: AI systems can adjust their behavior in real-time based on performance feedback
\end{itemize}

\paragraph{Scalability Properties}
\begin{itemize}
\item \textbf{Non-rivalry}: Unlike human career advancement, AI deployment is not subject to finite promotional opportunities
\item \textbf{Network effects}: AI system value often increases with usage, creating positive feedback loops
\item \textbf{Modular replication}: Successful AI components can be incorporated into multiple systems simultaneously
\end{itemize}

\paragraph{Measurement Precision}
\begin{itemize}
\item \textbf{Objective metrics}: AI performance is often more precisely quantifiable than human performance
\item \textbf{Real-time monitoring}: AI systems can provide continuous performance data streams
\item \textbf{Controlled experimentation}: AI capabilities can be tested under standardized conditions more easily than human abilities
\end{itemize}

\subsection{Fitness and Performance Operationalization for AI}

\subsubsection{AI Fitness Measures ($w_{ai}^g$)}
\begin{itemize}
\item Usage frequency and deployment rates across organizational tasks
\item Computational resource allocation (processing power, memory, storage)
\item Update frequency and development investment
\item User adoption rates and expansion to new domains
\item Integration depth into organizational workflows
\item Maintenance priority and technical support allocation
\end{itemize}

\subsubsection{AI Performance Measures ($\pi_{ai}^g$)}
\begin{itemize}
\item Accuracy rates and error frequencies in assigned tasks
\item Processing efficiency and speed of task completion
\item User satisfaction scores and interface effectiveness
\item Learning rate and adaptation to new scenarios
\item Output quality and consistency measures
\item Resource efficiency (output per unit of computational input)
\end{itemize}

\subsection{Cross-Term Interdependencies}

\subsubsection{Non-Independence of Selection Terms}
These three selection terms are not statistically independent:

\paragraph{Group-Individual Relationships}
Between-group selection depends partly on the distribution of within-group selection patterns. Groups with strong positive correlations between individual performance and fitness may exhibit different group-level dynamics than those with weak or negative correlations \cite{Traulsen2006}.

\paragraph{Human-AI Interactions}
Human and AI within-group selection influence each other through workplace dynamics:
\begin{itemize}
\item Successful humans may advocate for better AI tools
\item High-performing AI may enhance human productivity and career prospects
\item Competition between humans and AI for organizational resources may create negative correlations
\item Collaborative human-AI teams may generate positive spillover effects
\end{itemize}

\subsubsection{Institutional Modulation}
Ostrom's design principles can systematically influence these covariance structures:

\paragraph{Collective-Choice Arrangements}
May strengthen positive human performance-fitness correlations by ensuring that those affected by promotion and resource allocation decisions have input into the criteria and processes \cite{Ostrom2005}.

\paragraph{Monitoring Systems}
Can enhance the link between AI performance and continued deployment by providing transparent, accountable measures of system effectiveness \cite{March1991}.

\paragraph{Graduated Sanctions}
May create strong negative correlations between poor performance and fitness for both humans and AI by implementing escalating consequences for underperformance.

\subsubsection{Dynamic Evolution}
These covariance patterns evolve over time as:
\begin{itemize}
\item Organizations learn which performance metrics predict long-term success
\item Technological advances improve the measurability of both human and AI contributions
\item Competitive pressures shift the relative importance of different performance dimensions
\item Institutional rules adapt to better align individual and collective interests
\item Human-AI collaboration patterns mature and stabilize
\end{itemize}

This technical framework establishes the foundation for understanding how multi-level selection operates across heterogeneous agent types within organizations, providing the basis for analyzing the cross-agent interaction terms that follow.

\subsection{Formal Definition and Structure}

The cross-agent interaction terms are defined as:

\begin{equation}
\underbrace{\mathbb{E}_g\left[\text{Cov}_{(h, ai) \in g}(w_h^g, \pi_{ai}^g)\right] 
+ \mathbb{E}_g\left[\text{Cov}_{(h, ai) \in g}(w_{ai}^g, \pi_h^g)\right]}_{\text{cross-agent interaction terms}}
\label{eq:CrossAgentTerms}
\end{equation}

These consist of two distinct covariance terms that measure fundamentally different causal relationships. Unlike standard multi-level selection models that focus on within-type correlations \cite{Okasha2006, Wilson2007}, these cross-terms capture how the fitness of one agent type correlates with the performance of another agent type within the same organizational group.

\subsection{Individual Term Analysis}

\subsubsection{Term 1: Human Fitness and AI Performance}
The first term, $\mathbb{E}_g\left[\text{Cov}_{(h, ai) \in g}(w_h^g, \pi_{ai}^g)\right]$, measures the correlation between human fitness (reproductive success or organizational influence) and AI performance (task-specific output quality).

\paragraph{Positive Covariance Interpretation}
A positive covariance indicates that humans with higher fitness tend to work with better-performing AI systems. This pattern can arise through several mechanisms:

\begin{itemize}
\item \textbf{Selection effects}: More successful humans gain preferential access to superior AI tools and resources \cite{Brynjolfsson2017}
\item \textbf{Complementarity}: High-fitness humans possess superior skills for utilizing AI capabilities effectively \cite{Autor2015}
\item \textbf{Quality matching}: Organizations strategically pair their most valuable human talent with their best AI systems to maximize returns \cite{Aghion2017}
\item \textbf{Learning effects}: Successful humans provide better training, feedback, or oversight to AI systems \cite{hadfield2016cooperative}
\end{itemize}

\paragraph{Negative Covariance Interpretation}
Negative covariance suggests that high-fitness humans work with poorer-performing AI systems, which could indicate:

\begin{itemize}
\item \textbf{Substitution effects}: Highly capable humans rely less on AI assistance, potentially working with lower-quality systems \cite{Acemoglu2021harms}
\item \textbf{Resource scarcity}: Limited availability of high-quality AI forces strategic trade-offs in allocation
\item \textbf{Comparative advantage}: Organizations deploy AI as a substitute rather than complement for human capability
\end{itemize}

\subsubsection{Term 2: AI Fitness and Human Performance}
The second term, $\mathbb{E}_g\left[\text{Cov}_{(h, ai) \in g}(w_{ai}^g, \pi_h^g)\right]$, measures the correlation between AI fitness (computational success, deployment frequency, or resource allocation) and human performance (task-specific output quality).

\paragraph{Positive Covariance Interpretation}
Positive covariance indicates that more successful AI systems tend to work with better-performing humans:

\begin{itemize}
\item \textbf{Symbiotic matching}: High-performing AI thrives when paired with competent human operators who can effectively interpret and act on AI outputs \cite{Wilson2017}
\item \textbf{Feedback loops}: Better-performing humans provide higher-quality inputs, training data, or oversight that improve AI performance over time \cite{Russell2019}
\item \textbf{Strategic deployment}: Organizations deploy their most successful AI systems where human capability is highest to maximize synergies \cite{McAfee2017}
\item \textbf{Co-evolution}: Joint optimization processes where human-AI teams evolve together toward higher performance \cite{rahwan2019machine}
\end{itemize}

\paragraph{Negative Covariance Interpretation}
Negative covariance suggests that successful AI systems work with poorer-performing humans:

\begin{itemize}
\item \textbf{Compensatory deployment}: AI is strategically used to augment weak human performance rather than amplify strong performance
\item \textbf{Displacement effects}: Highly capable AI reduces performance requirements for human partners \cite{Brynjolfsson2014}
\item \textbf{Efficiency allocation}: Organizations use successful AI to raise the floor of human performance rather than lift the ceiling
\end{itemize}

\subsection{Computational Implementation}

To implement these covariance calculations empirically, researchers must address several technical challenges:

\subsubsection{Defining Interaction Pairs}
For each organizational group $g$, identify all human-AI pairs $(h, ai)$ that actually interact, rather than merely co-exist within the organization. This requires mapping:

\begin{itemize}
\item Direct collaboration relationships (e.g., human operators using specific AI tools)
\item Indirect dependencies (e.g., human decision-makers relying on AI-generated information)
\item Temporal interaction patterns (e.g., sequential workflows involving both humans and AI)
\end{itemize}

\subsubsection{Measurement Framework}
The covariance calculations require precise operationalization of four key variables:

\begin{itemize}
\item $w_h^g$: Human fitness measures (promotion rates, decision authority, resource allocation, peer evaluations)
\item $w_{ai}^g$: AI fitness measures (deployment frequency, computational resources allocated, update frequency, user adoption rates)
\item $\pi_h^g$: Human performance measures (output quality, accuracy, innovation metrics, customer satisfaction)
\item $\pi_{ai}^g$: AI performance measures (accuracy rates, efficiency gains, error rates, task completion times)
\end{itemize}

\subsubsection{Weighted Covariance Calculation}
The practical computation involves:

\begin{equation}
\text{Cov}_{(h, ai) \in g}(w_h^g, \pi_{ai}^g) := \frac{\sum_{h \in H_g} \sum_{ai \in A_g} \omega_{h,ai} (w_h^g - \overline{w}_h^g)(\pi_{ai}^g - \overline{\pi}_{ai}^g)}{\sum_{h \in H_g} \sum_{ai \in A_g} \omega_{h,ai}}
\label{eq:WeightedCovariance}
\end{equation}

In this context, $\omega_{h,ai}$ denotes the weight of interaction between human $h$ and AI agent $ai$, reflecting the strength or regularity of their cooperation, as described by \cite{Borgatti2009}.

\subsection{Theoretical Significance}

\subsubsection{Irreducibility of Cross-Terms}
These cross-agent interaction terms cannot be simplified or combined because they measure fundamentally different causal relationships:

\begin{itemize}
\item \textbf{Term 1}: How human evolutionary success affects access to or utilization of AI performance capabilities
\item \textbf{Term 2}: How AI evolutionary success affects association with or enhancement of human performance
\end{itemize}

This asymmetry reflects the different timescales and mechanisms through which humans and AI systems adapt and evolve within organizations \cite{March1991}.

\subsubsection{Emergent Organizational Properties}
The cross-covariation terms capture emergent organizational properties that pure within-type covariances miss:

\begin{itemize}
\item \textbf{Strategic complementarity}: The degree to which success of one agent type depends on quality of another \cite{Milgrom1990}
\item \textbf{Resource allocation patterns}: How organizations distribute AI capabilities among human workers and vice versa
\item \textbf{Evolutionary feedback}: How success of one agent type influences the development trajectory of the other
\item \textbf{Institutional effects}: How organizational rules and norms shape human-AI partnership patterns \cite{Ostrom2005}
\end{itemize}

\subsubsection{Biological Analogy}
In biological multi-level selection theory, analogous cross-terms would measure how the reproductive success of one species correlates with phenotypic traits of a symbiotic partner \cite{Frank1996}. Non-zero cross-covariances indicate that selection pressures are not independent across agent types—the evolutionary trajectory of humans shapes AI evolution and vice versa, creating complex co-evolutionary dynamics \cite{Thompson2005}.

\subsection{Implications for Institutional Design}

Understanding these cross-agent interaction terms has important implications for institutional design:

\subsubsection{Alignment Mechanisms}
Positive cross-covariances suggest successful alignment between human and AI agents, indicating that institutional rules effectively promote complementarity. Organizations should strengthen these mechanisms through:

\begin{itemize}
\item Incentive structures that reward human-AI collaboration
\item Training programs that enhance human-AI interface skills
\item AI development processes that incorporate human feedback loops
\end{itemize}

\subsubsection{Misalignment Detection}
Negative cross-covariances may signal institutional misalignment or suboptimal resource allocation. This suggests need for:

\begin{itemize}
\item Revised matching algorithms for human-AI partnerships
\item Reassessment of AI deployment strategies
\item Enhanced monitoring of human-AI interaction outcomes
\end{itemize}

\subsubsection{Dynamic Adaptation}
The magnitude and sign of these cross-terms can change over time as organizations learn and adapt. Institutions should incorporate mechanisms for:

\begin{itemize}
\item Regular assessment of human-AI interaction patterns
\item Adaptive resource allocation based on emerging complementarities
\item Evolution of governance structures to maintain positive cross-agent effects
\end{itemize}

This technical framework provides the foundation for empirical investigation of human-AI co-evolution within organizations and offers guidance for designing institutions that harness the full potential of hybrid human-AI teams.

\section{Comparison of Top-Down and Bottom-Up Institutional Rules}
\label{app:rule-comparison}

This appendix examines the trade-offs between top-down and bottom-up institutional rules within our multi-level selection framework.

\subsection{Contrasting Rule Types}

Top-down and bottom-up institutional rules represent fundamentally different approaches to establishing and maintaining cooperation within organizations:

\begin{itemize}
\item \textbf{Top-down rules} (e.g., the centralized sanctioning system in Section~\ref{sec:PD-Inst-eg}) are designed and enforced by a central authority. They provide organization-wide consistency and immediate implementation but require significant infrastructure and monitoring resources.

\item \textbf{Bottom-up rules} (e.g., tit-for-tat, win-stay-lose-shift strategies \cite{nowak1993strategy}) emerge through local interactions and are enforced through peer mechanisms. These decentralized approaches often have lower direct implementation costs but may diffuse more slowly through the organization.
\end{itemize}

\subsection{Economic Trade-offs in Implementation}

The fitness function $V_j = f_j(\pi^*_j) - c_j$ captures the critical trade-off between performance benefits and implementation costs. Bottom-up rules typically have different cost profiles than top-down approaches:

\begin{itemize}
\item \textbf{Lower monitoring costs}: Bottom-up rules often rely on direct agent observation rather than dedicated monitoring infrastructure

\item \textbf{Reduced enforcement overhead}: Sanctions are applied locally by peers rather than requiring centralized enforcement mechanisms

\item \textbf{Minimal coordination requirements}: Agents need only coordinate with direct interaction partners rather than alignment with organization-wide systems

\item \textbf{Lower administrative burden}: Bottom-up approaches typically require less formal documentation and bureaucratic maintenance
\end{itemize}

However, these cost advantages come with significant limitations that affect the performance component $f_j(\pi^*_j)$:

\begin{itemize}
\item \textbf{Constraint enforceability}: Bottom-up rules may lack reliable enforcement mechanisms for non-compliant agents, particularly in asymmetric power relationships

\item \textbf{Diffusion delays}: While top-down rules can be immediately implemented across an organization, bottom-up norms must spread through social learning and imitation, creating a temporal cost

\item \textbf{Coordination failures}: Without central orchestration, bottom-up rules may develop inconsistently across the organization, leading to incompatible norms between groups

\item \textbf{Equilibrium selection problems}: Multiple stable equilibria may emerge in different parts of the organization, creating inefficient norm fragmentation
\end{itemize}

\subsection{Evolutionary Dynamics of Rule Selection}

Within our replicator dynamics framework (Equation~\ref{eq:RuleReplicator}), we can model the competition between top-down and bottom-up rule configurations. Their relative success depends on organizational context:

\begin{itemize}
\item \textbf{High-trust environments} with stable agent relationships may favor bottom-up approaches, as the lower implementation costs outweigh the benefits of centralized enforcement

\item \textbf{Rapidly changing environments} may favor top-down rules that can be quickly modified and uniformly implemented across the organization

\item \textbf{Environments with high agent heterogeneity}—particularly those with mixed human and AI agents—may require hybrid approaches that combine centralized guidance with local adaptation
\end{itemize}

\subsection{Parallel Exploration of Rule Space}

One significant advantage of bottom-up rule evolution is the parallel exploration of rule space. When agents independently experiment with different norms and strategies, the organization simultaneously evaluates multiple potential institutional configurations. This creates an exploration-exploitation trade-off:

\begin{itemize}
\item Top-down rules exploit known effective mechanisms but explore limited institutional possibilities
\item Bottom-up approaches may discover novel, highly effective coordination mechanisms through a distributed search process
\end{itemize}

This parallel processing of rule evaluation can be particularly valuable in novel domains—such as human-AI coordination—where optimal institutional arrangements are not yet well understood \cite{Ostrom2005, Axelrod1997}.

\subsection{Hybrid Institutional Structures}

In practice, successful organizations often develop hybrid institutional structures that combine elements of both approaches:

\begin{itemize}
\item \textbf{Constitutional rules}: Top-down principles that establish boundaries and minimal standards for acceptable behavior

\item \textbf{Collective-choice rules}: Participatory mechanisms that allow those affected by rules to influence their design

\item \textbf{Operational rules}: Bottom-up norms that emerge within the constraints of higher-level rules but adapt to local conditions
\end{itemize}

This nested structure aligns with Ostrom's principle of nested enterprises \cite{Ostrom1990}, creating an institutional architecture that balances consistency with adaptability.

The evolutionary selection of these hybrid institutional structures represents a meta-level selection process, where not only specific rules but entire rule-generation mechanisms compete based on their ability to produce fitness-enhancing behavioral patterns within and between organizations.

\bibliographystyle{unsrt}
\bibliography{chapter_refs}

\end{document}